\documentclass[%
 reprint,
 amsmath,amssymb,
 aps,
]{revtex4-2}

\usepackage{graphicx}
\usepackage{dcolumn}
\usepackage{bm}
\begin{document}

\preprint{APS/123-QED}

\title{Machine learning for excitation energy transfer dynamics}

\author{Kimara Naicker}
 \email{kimaranaicker@gmail.com}
\author{Ilya Sinayskiy}
\author{Francesco Petruccione}
\affiliation{%
 Quantum Research Group, School of Chemistry and Physics, University of KwaZulu-Natal, Durban, KwaZulu-Natal, 4001, South Africa 
}%
\affiliation{%
 National Institute for Theoretical and Computational Sciences (NITheCS), South Africa
}%
\date{\today}

\begin{abstract}
A well-known approach to describe the dynamics of an open quantum system is to compute the master equation evolving the reduced density matrix of the system. This approach plays an important role in describing excitation transfer through photosynthetic light harvesting complexes (LHCs). The hierarchical equations of motion (HEOM) was adapted by Ishizaki and Fleming (J. Chem. Phys., 2009) to simulate open quantum dynamics in the biological regime. We generate a set of time dependent observables that depict the coherent propagation of electronic excitations through the LHCs by solving the HEOM. We solve the inverse problem using classical machine learning (ML) models as this is a computationally intractable problem. The objective here is to determine whether a trained ML model can perform Hamiltonian tomography by using the time dependence of the observables as inputs. We demonstrate the capability of convolutional neural networks to tackle this research problem. The models developed here can predict Hamiltonian parameters such as excited state energies and inter-site couplings of a system up to 99.28\% accuracy and mean-squared error as low as 0.65. 

\end{abstract}

\maketitle

\section{\label{sec:level1}Introduction}
During the first step of photosynthesis in light harvesting complexes, photons are absorbed by the antenna. While part of the energy of the photons is converted into heat in the form of molecular vibrations, most of the energy is captured as excitons which are subsequently transferred via chromophores to the reaction centre through a process labeled as excitation energy transfer (EET). It is at the reaction centre where photochemical reactions are triggered \cite{fenna}. 

The evidence for quantum coherence, which has no classical analogue, in the exciton transport process became undeniable in 2007 when Engel \textit{et al.} \cite{Engel07} used two-dimensional spectroscopic signatures \cite{Fleming06} to demonstrate quantum ‘beating’ within a photosynthetic complex at 77K, a result that was later confirmed at room temperature. Quantum beating in spectroscopic measurements provides a direct measure of quantum coherence on the appropriate energy and time scales \cite{Engel11}. 
While electronic coherence was first proposed as the source for the observed long-lived quantum coherence \cite{Engel07}, experimental and theoretical evidence has also supported proposals that the phenomenon resulted from a mixture of electronic and vibrational states which is referred to as `vibronic' coherence \cite{fuller}. 

The idea of quantum coherence playing a role in photosynthesis arose from observations that some energy or electron transfer processes in bacterial and plant pigment–protein complexes are efficient to an extent that exceeds explanation using only classical theory. Engel \textit{et al.} \cite{Engel07} investigated photosynthetic EET in the Fenna–Matthews–Olson (FMO) protein of green sulfur bacteria \cite{fenna, li}. The FMO complex was the first chlorophyll-containing protein that was crystallized \cite{Olson04}. Prior to mass spectrometry measurements of the protein that confirmed the existence of an eighth pigment, it was accepted that the protein consisted of seven pigments \cite{Wen11}. Subsequently, FMO is made up of three subunits each consisting of eight bacteriochlorophyll molecules. The FMO complex serves as a bridging energy wire as it is tasked with transporting energy in the form of light harvested in the antenna chlorosome to the reaction centre pigments \cite{fenna}. It represents an important model in EET and has been extensively studied experimentally and theoretically. Engel and collaborators succeeded in observing long-lasting quantum effects providing direct evidence for long-lived electronic coherence \cite{Engel11}. The observed coherence lasts for time scales similar to the EET timescales, implying that electronic excitations move coherently through the FMO protein rather than by previously proposed incoherent hopping motion \cite{Vulto99, Grondelle00}. Panitchayangkoon \textit{et al.} \cite{Panitchayangkoon} presented evidence that quantum coherence survives in FMO at physiological temperature for at least 300 fs which is long enough to impact biological energy transport. Collini \textit{et al.} \cite{Collini} made observations that provide evidence for quantum coherent sharing of electronic excitation across proteins under biologically relevant conditions. They suggest that distant molecules within the photosynthetic proteins are ‘wired’ together by quantum coherence for more efficient light-harvesting in cryptophyte marine algae. Lee \textit{et al.} \cite{Lee07} present experimental results in which they suggest that correlated protein environments preserve electronic coherence in photosynthetic complexes and allow the excitation to move coherently in space which enables highly efficient energy harvesting and trapping in photosynthesis. However, explanations for observed long-lived coherence have evolved during the past decade \cite{dattani, Dattani12, Duan17}. More recently, Fuller \textit{et al.} \cite{fuller} and Thyrhaug \textit{et al.} \cite{Thyrhaug18} suggest that the coherences are of a mixed electronic-vibrational nature and may enhance the rate of charge separation in oxygenic photosynthesis.

Understanding the relationship between the structure of light harvesting complexes and their excitation energy transfer dynamics is of importance in many applications. Insight into long-lived quantum coherence in EET processes can be gained through the reduced equation of motion and the numerically exact formalism of quantum dynamics adopted to study EET processes is the Hierarchical Equations of Motion (HEOM) derived by Tanimura and Kubo \cite{Tanimura89}. 

Machine learning is a well established tool that has been actively applied in various ways to address physical problems \cite{Carleo19}. One common strategy is to use supervised learning in which an algorithm is trained with datasets that are labeled beforehand then, the goal of the algorithm is to establish a general rule for assigning labels to data outside the training set. This approach can be used to identify distinct phases of matter and the transitions between them, one of the central problems in condensed-matter physics, that has been tackled by Carrasquilla and Melko \cite{Carrasquilla17}. Machine learning techniques have been used to represent and solve quantum systems such as in a work by Carleo and Troyer \cite{Carleo17} where the authors introduce an ansatz capable of both finding the ground state and describing the unitary time evolution of complex interacting quantum systems. More specifically and in a bid to investigate open quantum systems further, the applications of supervised machine learning that we are interested in are related to forms of approximating solutions to open quantum system dynamics such as where multi-layer perceptrons have been used to obtain the exciton dynamics of large photosynthetic complexes \cite{Hase16} and to better understand the relationship between the structure of light-harvesting systems and their excitation energy transfer properties \cite{guzik1}; where recurrent neural networks were used to model quantum systems interacting with an unknown environment \cite{Banchi18} and where convolutional neural networks were used to predict long-time dynamics of an open quantum system \cite{Rodriguez19}. 

\begin{figure}
    \includegraphics[width=\linewidth]{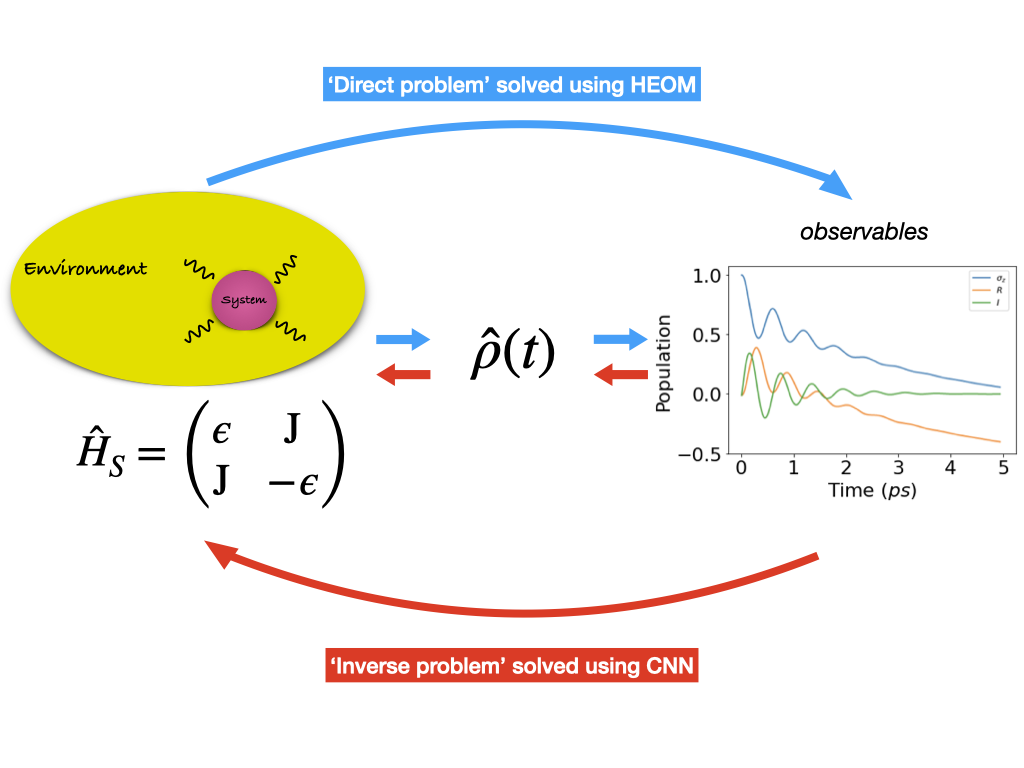}
    \caption{Machine learning Hamiltonian tomography in open quantum systems. The `direct problem' is the process of using the Hamiltonian of an open quantum system $\hat{H}_{S}$ as an input to the HEOM to calculate the evolution of the reduced density matrix $\hat{\rho}(t)$ to produce the time dependent observables. The `inverse problem' addressed in this study is the process of using the observables as an input to the CNN to reproduce the system Hamiltonian.}
    \label{fig:fig0}
\end{figure}

The primary focus of this work is on using classical machine learning models to study the quantum dynamics of EET. We can generate a time dependent set of observables that depict the coherent movement of electronic excitations through a photosynthetic pigment-protein complex by solving the HEOM. Here we develop a scalable and efficient tool for the description of the dynamical properties of open quantum systems by use of a trained convolutional neural network (CNN) to solve the inverse problem. This means that the objective is to determine whether a trained CNN can accurately describe the system under study, by predicting the parameters of the system Hamiltonian such as excited state energies and inter-site couplings, when given this time dependent data of varying length (see Figure~\ref{fig:fig0}). 

In Section~\ref{sec:1}, we describe the HEOM framework, Section~\ref{sec:2} is used to introduce and explain the use of machine learning in this context. In Section~\ref{sec:results}, we present the numerical results which show that our models can successfully reproduce the Hamiltonian of a system and that our method can be easily adapted to study more complex systems. Finally, Section~\ref{sec:conclusions} is devoted to concluding remarks.

\section{\label{sec:1}Hierarchical Equations of Motion}
One of the viable approaches to explore long-lived quantum coherence and its interplay with the protein environment in EET processes is through the reduced equation of motion. In this approach, the key quantity of interest is the reduced density matrix, i.e., the partial trace of the total density matrix over the environmental degrees of freedom \cite{BRE02}. 

Typical situations in photosynthetic EET are such that the electronic coupling strengths, between chromophores and their local environment phonons, span a similar range as the reorganization energies, which characterize the time scale of the coupled phonons relaxing to their respective equilibrium states \cite{Banchi18, Strumpfer12}. However, these site-dependent reorganization processes cannot be described by theories that rely on the Markov approximation as it requires the phonons to relax to their equilibrium states instantaneously, that is, the phonons are always in equilibrium even under the electron-phonon interaction.

In order to go beyond the Markov approximation, Tanimura and Kubo \cite{Tanimura89} developed a new theoretical framework, the HEOM, which can describe the site-dependent reorganization dynamics of environmental phonons. Ishizaki and Fleming \cite{ishizaki1} adapted this formalism to suit the quantum biological regime which is the form we employ. HEOM is a numerically exact method which accurately accounts for the reorganization process in which the vibrational coordinates rearrange to their new equilibrium positions upon electronic transition from the ground to the excited potential energy surface. It can describe quantum coherent wavelike motion and incoherent hopping in the same framework and reduces to the conventional Redfield \cite{Redfield57, Redfield65, Pollard96} and Förster \cite{Forster48, Forster65} theories in their respective limits of validity. In the following Subsection~\ref{subsec:Theory}, we highlight the theory required to describe EET dynamics in a photosynthetic complex \cite{ishizaki2, Ishizaki10}.

\subsection{\label{subsec:Theory}Theory}
The total Hamiltonian is composed of the Hamiltonian of the system, bath and system-bath interaction,
\begin{equation}
\label{eqn:Htot}
    \hat{H}_{TOT} = \hat{H}_{S} +\hat{H}_{B}+\hat{H}_{SB} .
\end{equation}

The Hamiltonian of the system refers to the electronic states of a complex containing N pigments,
\begin{equation}
\label{eqn:Hs}
    \hat{H}_{S} = \sum_{j=1}^{N} |j\rangle\epsilon_j\langle j| + \sum_{k\neq j}^{N} |j\rangle J_{jk}\langle j| ,
\end{equation}
where $\epsilon_{j}$ is the excited state energy of the $j$th site and $J_{jk}$ denotes the electronic coupling between the $j$th and $k$th sites.

Here we consider that each pigment is coupled to a separate bath. The bath Hamiltonian represents the environmental phonons, 
\begin{equation}
    \hat{H}_{B} = \sum_{j=1}^N \hat{H}_{B_{j}}, \quad \hat{H}_{B_{j}} = \sum_{k} \hbar \omega_{j,k} \bigg(\frac{\hat{p}_{j,k}^2 + \hat{q}_{j,k}^2}{2}\bigg) ,
\end{equation}
where $p$ is the conjugate momentum, $q$ is the dimensionless coordinate and $\omega_{j,k}$ is the frequency of the $j$th site and $k$th phonon mode, respectively. The last term of Eq.~\eqref{eqn:Htot} represents the fluctuations in the site energies caused by the phonon dynamics,
\begin{equation}
    \hat{H}_{SB} = \sum_{j=1}^{N} \hat{u}_j |j\rangle\ \langle j|  , \quad \hat{u}_j = \sum_{k} g_{j,k}\hat{q}_{j,k} ,
\end{equation}
where $g_{j,k}$ is the coupling constant between the $j$th site and $k$th phonon mode.

The spectral density $\mathcal{J}_{j}(\omega)$ specifies the coupling of an electronic transition of the $j$th pigment to the environmental phonons through the reorganization energy $\lambda_{j}$ and the timescale of the phonon relaxation $\gamma_{j}$. Here it is expressed as the Ohmic spectral density with Lorentz-Drude cut-off, $\mathcal{J}_{j}(\omega) = 2\lambda_{j} \gamma_{j} \omega / (\omega^2 + \gamma_{j}^2)$. 

We focus on the application of this theory to EET at physiological temperatures of around 300 K. Hence, when the high-temperature condition characterized by $\hbar \gamma_{j} / k_{B}T << 1$ is imposed, the following hierarchically coupled equations of motion is obtained \cite{ishizaki1},
\begin{eqnarray*}
    \frac{\partial}{\partial t} \hat{\sigma}(\textbf n, t) = - \left( i \hat{\mathcal{L}_e} + \sum_{j=1}^{N} n_j\gamma_j \right) \hat{\sigma}(\textbf n, t) \nonumber \\
\end{eqnarray*}
\begin{eqnarray}
\label{eqn:HEOM}
    + \sum_{j=1}^{N} \bigg[ \hat{\Phi}_j\hat{\sigma}(\textbf n_{j+}, t) + n_j\hat{\Theta}_j\hat{\sigma}(\textbf n_{j-}, t) \bigg].
\end{eqnarray}

In Eq.~\eqref{eqn:HEOM}, $\textbf{n} \equiv (n_{1}, n_{2},\dots, n_{N})$ for sets of non-negative integers and $\textbf n_{j\pm}$ differs from $\textbf{n}$ by changing the corresponding $n_{j}$ to $n_{j} \pm 1$. Furthermore in Eq.~\eqref{eqn:HEOM}, the element $\hat{\sigma}(\textbf 0, t)$ is identical to the reduced density operator $\hat{\rho}(t)$, while the rest are auxiliary density operators. Moreover, the Liouvillian corresponding to the Hamiltonian $\hat{H}_{S}$ is denoted by $\hat{\mathcal{L}_e}$ and the relaxation operators $\hat{\Phi}_{j}$ and and $\hat{\Theta}_{j}$ are given by Eqs.~\eqref{eqn:phi} and \eqref{eqn:theta},
\begin{equation}
\label{eqn:phi}
    \hat{\Phi}_j = i V_j^\times , \quad V_j^\times y = [V_j, y], 
\end{equation}
\begin{equation}
\label{eqn:theta}
    \hat{\Theta}_j = i \bigg( \frac {2\lambda_j}{\beta \hbar^2} V_j^\times -i\frac {\lambda_j}{\hbar} \gamma_j V_j^\circ \bigg) , \quad V_j^\circ y = \{V_j, y\} .
\end{equation}

Formally the hierarchy in Eq.~\eqref{eqn:HEOM} is infinite and cannot be numerically integrated. In order to make this problem tractable, the hierarchy can be terminated at a certain depth. There are several methods of doing so and in this work we have chosen the following termination condition following Ishizaki and Fleming \cite{ishizaki2}. For the integers $\textbf n = (n_1, n_2,\dots, n_N)$ and for characteristic frequency $\omega_{e}$ of $\hat{\mathcal{L}_e}$ where
\begin{equation}
\label{eqn:terminator}
    \mathcal{N} \equiv \sum_{j=1}^{N} n_{j} \gg \frac{\omega_{e}}{\mathrm{min}(\gamma_1, \gamma_2,\dots, \gamma_N)} ,
\end{equation} 
Eq.~\eqref{eqn:HEOM} is replaced by
\begin{eqnarray*}
    \frac{\partial}{\partial t} \hat{\sigma}(\textbf n, t) = -i \hat{\mathcal{L}_e} \hat{\sigma}(\textbf n, t) .
\end{eqnarray*}

\section{\label{sec:2}Machine Learning Approach}
There exist numerous studies \cite{Hase16, guzik1,Banchi18, Rodriguez19,Montavon13,Gomes17} of machine learning techniques applied to accelerate computations by many orders of magnitude at a reasonable level of accuracy. A machine learning model can be leveraged to predict the reduced density matrix of a system given the Hamiltonian of the system. In this approach, the model is trained and tested on a large and diverse enough dataset of Hamiltonians and corresponding reduced density matrices such that it may learn patterns in the data and be able to present highly accurate predictions without having any knowledge of the theory or in this case, HEOM. 

However, in an experimental setting one may gather certain time-dependent observational data and subsequently need to use these findings to attain the Hamiltonian of the excitonic system under study. The use of machine learning models in this work is to act as a \textit{blackbox} which one can input excitation energy transfer observations into and obtain Hamiltonian parameters from.

In the subsequent sections, Subsection~\ref{subsec:2.1} we describe the machine learning basics required to follow the study. Thereafter, in Subsection~\ref{subsec:2.2} we generate multiple datasets comprising of excited state population dynamics and corresponding Hamiltonian parameters and in Subsection~\ref{subsec:2.3} we design the supervised machine learning model architecture to be used for making predictions based on the generated datasets.  

\subsection{\label{subsec:2.1}Supervised machine learning}
Machine learning algorithms are used to learn underlying patterns embedded in the data. In the realm of classical machine learning, there exist three broad types classified by the amount and type of supervision models get during training: supervised, unsupervised and reinforcement learning. In supervised learning, the training set fed to the model includes the true solutions called labels \cite{Aurelien19, Murphy12}.   

A neural network is a machine learning model whose structure is inspired by the networks of biological neurons found in our brains \cite{McCulloch43}. They are made up of layers of neurons which are core processing units of the network. Usually these layers contain an input layer to receive input features, an output layer to make final predictions and hidden layers which perform most of the computations done by the network. Convolutional neural networks (CNNs) \cite{LeCun98} is a class of neural networks which emerged from the study of the brain’s visual cortex \cite{Hubel59, Fukushima80}. CNNs specialize in processing data that has a grid-like topology. The human brain processes information when we see an image as each neuron works in its own restricted region of the visual field called the receptive field and is connected to other neurons in a way that covers the entire visual field. Just as each neuron responds to stimuli only in its receptive field in the biological vision system, each neuron in a CNN processes data only in its receptive field. The layers are arranged such that they detect simpler patterns first and more complex patterns deeper into the network. A rich description of how the convolution operation works and the advantages of using CNNs has been written by Goodfellow \textit{et al.} \cite{Goodfellow16}.

In this work, machine learning has been leveraged to solve the HEOM \textit{inversely} without explicitly solving the equations at all so that predictions of the parameters of the Hamiltonian of a system can be made when given time dependent observations.

\subsection{\label{subsec:2.2}Generating the database}
To demonstrate the capabilities of our machine learning approaches for the regression task at hand, we investigate three datasets of increasing complexity that are randomly generated excitonic systems. The parameters of the Hamiltonians in these datasets are motivated by and sampled around the same order of magnitude as those that are typical of the light-harvesting pigment-protein FMO. 

We impose a linear chain such that only neighbour-neighbour couplings are permitted and for simplicity, we consider that the transition rates are strictly real in value. When sampling the Hamiltonian parameters we consider the excited state energy $\epsilon_{j}$ and inter-site coupling $J_{jk}$ with respect to $\epsilon_{1}$. Hence, the Hamiltonian of an N-level system would require $2(N-1)$ real parameters. In constructing the first dataset, we consider two-level excitonic systems which allow for excitation energy transfer between two excited states. In this case, there are two Hamiltonian parameters needed to describe the two-level system as seen in Eq.~\eqref{eqn:Hs} where $N=2$. Therefore, by solving the HEOM we obtain the time evolution of an N-dimensional reduced density matrix for each sample. In a similar way, the second and third datasets consider three-level and four-level systems which require four and six Hamiltonian parameters, respectively.  
\begin{figure}[h]
    \centering
    \includegraphics[scale=0.63]{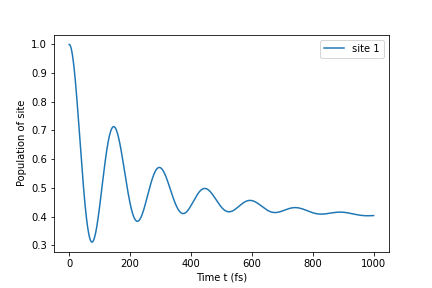}
    \caption{Time evolution of the population of site 1 of a 2-level system calculated by the HEOM, Eq.~\eqref{eqn:HEOM}. This calculation was done at T = 300 K where the reorganization energy and the phonon relaxation time are set to be $\lambda$ = 35 $cm^{-1}$ and $\gamma$ = 106.1767 $cm^{-1}$, respectively. Other parameters were fixed to be E = 100 $cm^{-1}$ and $J_{12}$ = 100 $cm^{-1}$.}
    \label{fig:fig1}
\end{figure}
The following data is captured and stored for each sample in each dataset: Hamiltonian parameters which are used as labels for the machine learning model and, the time evolution of each element of the reduced density matrix to be used as input features for the machine learning model. An example of an input feature for a two-level system can be seen in Figure~\ref{fig:fig1} which depicts the time evolution of the population of site 1 which corresponds to the first element in the reduced density matrix. For each dataset, 25000 Hamiltonians were generated by sampling a uniform distribution for excited state energies and inter-site couplings within a fixed range of values around those that are typical of FMO complex shown in Table~\ref{tab:table1}. Only neighbour-neighbour couplings were considered to be non-zero, hence, $J_{jk}$ were only sampled for each site's nearest neighbours. Furthermore, the values of excited state energy $\epsilon_{j}$ have been sampled with respect to 12400 $cm^{-1}$ for all sites \cite{guzik1, Caruso09}.
\begin{table}[h]%The best place to locate the table environment is directly after its first reference in text
\caption{\label{tab:table1}%
Lower and upper limits in between which excited state energies $\epsilon_{j}$ and inter-site couplings $J_{jk}$ for each site's nearest neighbours were uniformly sampled to generate the three datasets of this study. 
}

\begin{ruledtabular}
\begin{tabular}{ c c c }
\textrm{Dataset}&\textrm{$\epsilon_{j} [cm^{-1}]$}&\textrm{$J_{jk} [cm^{-1}]$}
\\ 
\colrule
2 & [-100, 100] & [-100, 100] \\
3 & [-100, 100] & [-100, 100] \\
4 & [-100, 100] & [-100, 100] \\
\end{tabular}
\end{ruledtabular}
\end{table}

By explicitly solving the HEOM, we compute the time evolution of the reduced density matrix for each Hamiltonian in each of our datasets. As opposed to the traditional 4th-order Runge-Kutta method, the numerical propagation was done by an exponential expansion method described by Dattani \textit{et al.} for 1 ps represented by 5000 time steps \cite{dattani}. Each computation took approximately 1 s, 1.5 s and 2 s for a two-level, three-level and four-level system sample, respectively. For all samples in all datasets, the HEOM were truncated at a depth of 3 (refer to Eq.~\eqref{eqn:terminator}). For all Hamiltonians we assumed identical Drude–Lorentz spectral densities describing the influence of the bath on each excited state. For all datasets, simulations were done at temperature T = 300 K, reorganization energy $\lambda$ = 35 $cm^{-1}$ and phonon relaxation time $\gamma$ = 106.1767 $cm^{-1}$ \cite{ishizaki1}. Python’s main scientific libraries used are NumPy, pandas, and Matplotlib and Python frameworks for machine learning tasks are Scikit-Learn, TensorFlow and Keras.

\subsection{\label{subsec:2.3}Model architecture}
The architecture of our CNNs are designed for supervised learning of excitation energy transfer dynamics. As mentioned in Subsection~\ref{subsec:2.2}, the elements of the reduced density matrices obtained by solving the HEOM are the inputs known as features for the CNN. To reduce the computational cost of model training and to improve the performance of the model, a feature selection process was carried out to reduce the number of input variables. During the process of optimization of our CNNs it was found that rather than the entire reduced density matrix, only the diagonal elements and their nearest-neighbours were required as input features to allow the models to perform best. Data scaling was performed to transform all datasets as a pre-processing step by fitting a scaling object only to the training data then using it to tranform the training/ validation and test sets. All features of all datasets were normalized by rescaling the data into the range [0,1]. Similarly, the labels were transformed so as to be normally distributed such that the mean of the values is 0 and the standard deviation is 1. For each dataset, all features were reshaped into 4-dimensional tensors (number of samples, number of time steps, number of features, 1) and provided as input features to the CNNs, which were used to predict excited state energies and inter-site couplings known as labels. Since the input features of neural networks need to be of fixed size, we construct separate CNNs for each dataset in order to treat the different dimensionalities of the Hamiltonians. These CNNs only differ by their input and output shapes. The hidden layers of the CNNs consist of two 2D convolutional layers with by a max pooling operation layer, the previous three layers are repeated and then followed by a flattening layer and three dense layers. Goodfellow \textit{et al.} \cite{Goodfellow16} describe general guidelines for choosing which architectures to use in which circumstances. The full architecture of the models with hyperparameters can be found in Section~\ref{sec:appendix}.

The 25000 Hamiltonians of each dataset were split into three sets: a training set of $85\%$ of all Hamiltonians where $80\%$ of these are used for training CNN model instances with particular hyperparameters and the other $20\%$ form a validation set used to evaluate the CNN architecture during optimization of the hyperparameters and a test set of $15\%$ of all Hamiltonians to probe out-of-sample prediction accuracies. Noteworthy, after splitting the data into train, validation and test sets, the distribution of the Hamiltonian parameters (labels) of each of the subsets maintained their uniform distributions. All constructed CNN models were trained with 100 data points per batch and the ADAM optimizer with a learning rate of 0.001 until the mean squared error (MSE) on the validation set increased over three full consecutive training epochs on a computational cluster. The MSE provides a direct quantitative check of how close estimates or forecasts are to actual values by taking the difference between them and squaring them. This measure works well in ensuring that our trained model has no outlier predictions with large errors since it allocates larger weight to theses errors due to the squaring part of the function. Although the MSE was used to evaluate the cost function throughout the training and testing processes, the coefficient of determination, $R^{2} = 1 - \frac{\sum_{j} (y_{i} - \overline{y})^2}{\sum_{j} (y_{i} - f_{i})^2}$, was calculated as an alternate accuracy measure. $R^{2}$ depends on the ratio of total deviation of results described by the model; it can be interpreted as the percentage of variation in the dependent output attribute that the model is capable of explaining. Both of these measures were used in analyzing the results in order to determine the predictive power of the model. All CNN models were generated and trained using the Tensorflow package.

\section{\label{sec:results}Results}
In this section, we demonstrate the capabilities of our trained CNN models by analyzing the MSE between predicted Hamiltonian parameters and those used in the numerically exact HEOM calculations and, the coefficient of determination ($R^2$ score). Our trained models predict Hamiltonian parameters for test (out-of-sample) data at almost the same accuracy as for training and validation data on which CNN model parameters and hyperparameters were optimized. This demonstrates the ability of our models to generalize well to previously unseen data and to provide out-of-sample predictions with high accuracy. To support this conclusion, we present the results of the 5-fold cross-validation of each of our models in Table~\ref{tab:table2} which was carried out prior to training. A complete set of samples is randomly shuffled and split into the specified number of folds to form smaller sample groups. Cross-validation is a re-sampling procedure then used to evaluate the performance of a machine learning ansatz on each of the limited data sample sets formed. The model is then fit using the K-1 folds and validated using the remaining Kth fold. This process is repeated until every K-fold has served as a test set then the average of the recorded scores are captured.
\begin{table}[h]
\caption{\label{tab:table2} K-fold cross-validation results for the CNN model used for each dataset where K=5.}
\begin{ruledtabular}
\begin{tabular}{c | c | c | c | c}
Dataset&\textrm{MSE}&\textrm{MSE error}&\textrm{$R^2$ score}&\textrm{$R^2$ score error}
\\ 
\colrule
2 & 0.01 & 0.01 & 99.94 & 0.01 \\
3 & 0.06 & 0.01 & 94.73 & 0.01 \\
4 & 0.10 & 0.01 & 91.63 & 0.02 \\
\end{tabular}
\end{ruledtabular}
\end{table}

Table~\ref{tab:table3} summarizes the results for the predicted Hamiltonian parameters for our three generated datasets where the full length of 1 ps equating to 5000 timesteps was considered for all selected input features. 
\begin{table}[h]
\caption{\label{tab:table3} Mean squared error (MSE) and coefficient of determination ($R^2$ score) of Hamiltonian parameters used in HEOM calculations and predicted by the trained CNNs. For all three datasets, the full time length of 1 ps for all features were input to the model. The results of the training, validation and test sets are shown, separately.}
\begin{ruledtabular}
\begin{tabular}{c | c c | c c | c c}
Dataset& \multicolumn{2}{c}{Train} &\multicolumn{2}{c}{Validation} &\multicolumn{2}{c}{Test}\\
&\textrm{MSE}&\textrm{$R^2$ score}&\textrm{MSE}&\textrm{$R^2$ score}&\textrm{MSE}&\textrm{$R^2$ score}
\\ 
\colrule
2 & 0.83 & 99.16 & 0.65 & 99.34 & 0.70 & 99.28 \\
3 & 2.92 & 97.06 & 2.86 & 97.11 & 3.28 & 96.64 \\
4 & 6.58 & 93.42 & 6.91 & 93.04 & 7.31 & 92.63 \\
\end{tabular}
\end{ruledtabular}
\end{table}
Furthermore, we can deduce that the architectures of the neural networks are well-balanced and neither in the regime of over- or under- fitting which would result in a large discrepancy in errors between the training and validation datasets. The predictions carried out with the CNN architectures only show variation in their performances depending on the dataset under study. Overall we find a high accuracy of our predictions and small mean squared errors on the datasets which are in the range between 0.70 for a 2-level system and 7.25 for the largest considered 4-level system. The 4-level system dataset exhibits the most diverse transfer properties which explains the larger mean squared errors in the predictions when compared to the other datasets.

\begin{figure}
    \includegraphics[width=\linewidth]{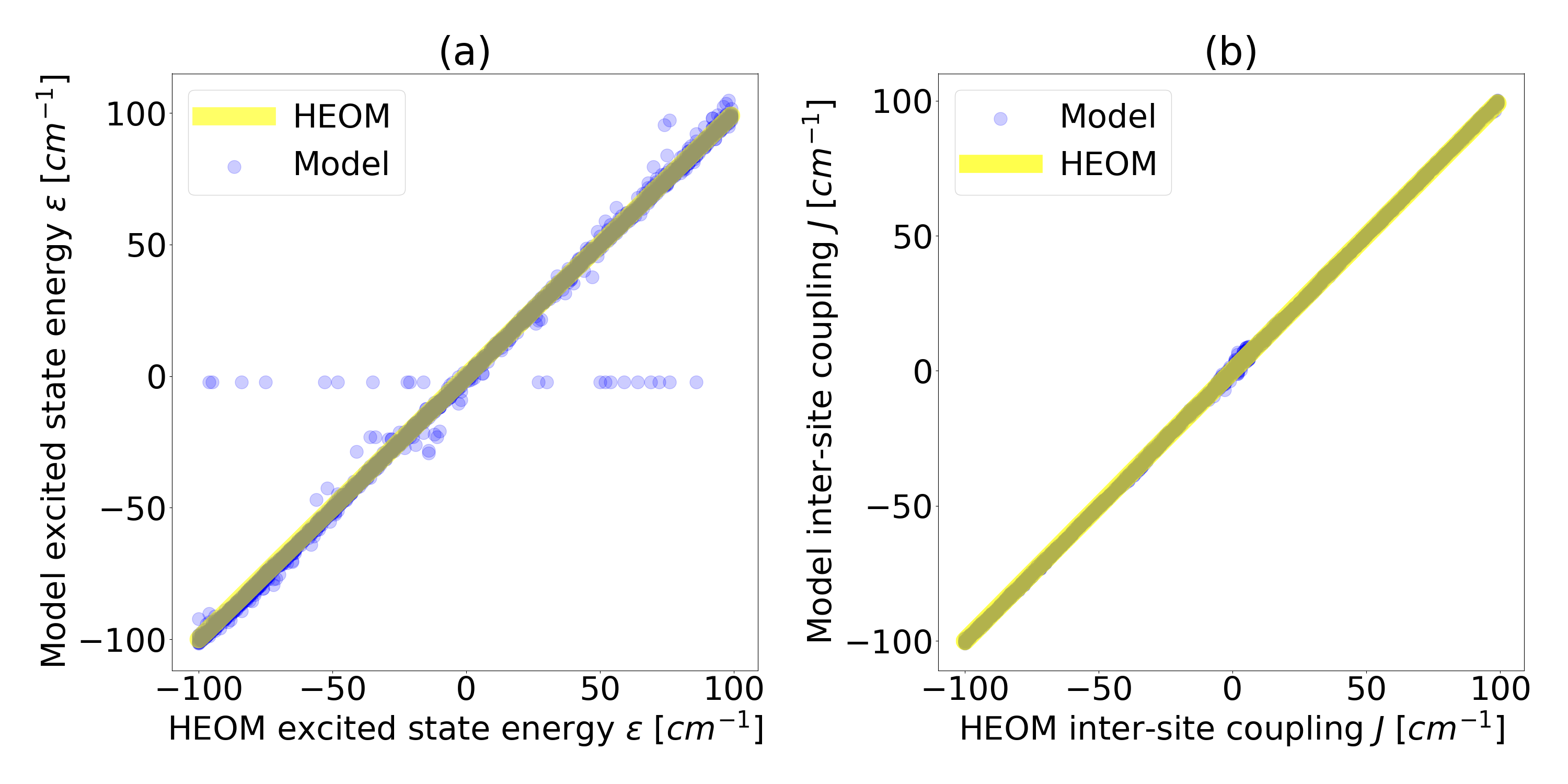}
    \caption{(a) Excited electronic energy and (b) electronic coupling as computed with the HEOM approach compared to prediction from CNN model for a two-level system. The blue dots represent the model predictions. The yellow line indicates perfect agreement between HEOM results and predictions by the model.}
    \label{fig:fig2}
\end{figure}

Figure~\ref{fig:fig2} provides a visual comparison of Hamiltonian parameters as computed with the HEOM approach and predictions with trained CNN models. In Figure~\ref{fig:fig2}(a), one can observe that there exist a group of points of excited state energies which are not well predicted by the model. As random Hamiltonians were generated in the data preparation step, the Hamiltonians that correspond to these points cover the full range of allowed excited state energies as seen in Figure~\ref{fig:fig2}(a), however, all of these points also correspond to the range of inter-site couplings that is [-30, 30] $cm^{-1}$. This is known as the overdamped regime where the time dependent observables do not display a coherent but rather a purely dissipative behaviour, hence, the CNN does not perform well in differentiating between these represented systems. 

As highlighted in section~\ref{subsec:2.2}, feature selection is critical to the performance of trained models. The time evolution of selected elements of the reduced density matrices used as input features each contain data for 1 ps long. The next objective of this work was to determine by how much this time series data could be shortened by while still maintaining the accuracy measures achieved with the full dataset. The results obtained are shown in Table~\ref{tab:table4}.
\begin{table}[h]
\caption{\label{tab:table4} Mean squared error (MSE) and coefficient of determination ($R^2$ score) of Hamiltonian parameters used in HEOM calculations and predicted by the trained CNNs. For each dataset, the time length in fs for all features which were input to the model are given. The results of the training, validation and test sets are shown, separately.}
\begin{ruledtabular}
\begin{tabular}{c|c|cc|cc|cc}
Dataset& Time & \multicolumn{2}{c}{Train} &\multicolumn{2}{c}{Validation} &\multicolumn{2}{c}{Test}\\
&&\textrm{MSE}&\textrm{$R^2$ score}&\textrm{MSE}&\textrm{$R^2$ score}&\textrm{MSE}&\textrm{$R^2$ score}
\\ 
\colrule
2 & 400 & 0.77 & 99.22 & 0.95 & 99.04 & 0.96 & 99.03\\
3 & 400 & 2.39 & 97.59 & 3.01 & 96.95 & 3.02 & 96.91\\
4 & 400 & 6.62 & 93.36 & 7.82 & 92.08 & 7.27 & 92.70\\
\end{tabular}
\end{ruledtabular}
\end{table}
From this we can deduce that the full time length is not required to maintain high accuracy, rather only 400 fs are sufficient for 2, 3 and 4 level systems. The observed prediction errors are also consistent with the complexity of the system dynamics for each of the three datasets which indicates that CNN models generally benefit from a wider sampling of the input parameter space.
%test on ishizaki data for 2 level. from a finer sampling of the input parameter space.

\section{\label{sec:conclusions}Conclusions}
During photosynthesis in light harvesting complexes, energy is transferred from antenna pigments to the reaction center to trigger photochemical reactions. The formalism adopted to study excitation energy transfer (EET) processes is the Hierarchical Equations of Motion (HEOM). This work focuses on leveraging classical machine learning models to study the dynamics of EET within open quantum systems. We propose the use of a trained convolutional neural network (CNN) to perform Hamiltonian tomography when given input data that represents the dynamics of EET through the open quantum system over time.
We have discussed the investigation of EET for 2-, 3- and 4-level systems where linear chain configurations were imposed. The performance of the models were gauged by mean-squared error and coefficient of determination measures.
We have proven the capabilities of CNNs and supervised machine learning as an efficient tool for solving the inverse problem of the HEOM by employing a model to predict the parameters of Hamiltonian's when given underlying time dependent observations as features. In particular, we have shown that using a trained CNN one can predict the Hamiltonian parameters such as excited state energy and electronic coupling up to $99.28\%$ accuracy and mean-squared error as low as 0.65. We propose the use of the trained CNNs as an efficient way to study the excitation energy transfer dynamics of biological complexes. 
An improvement that can be investigated in future is a more sophisticated algorithm that will be able to distinguish between the systems in the overdamped regime that can be described by the set of observables that are purely dissipative. This work can be developed further in a few ways, either by investigating fully connected neural networks or by developing a model such that it may be independent of the dimension of input data. The latter modification would allow the user to input data such that in return the model may determine the dimension of the system under study which represents the number of pigments in the complex.

\section{\label{sec:acknowledgement}Acknowledgments}

This work is based upon research supported by the South African Research Chair Initiative, Grant No. 64812 of the Department of Science and Innovation and the National Research Foundation of the Republic of South Africa. Support from the NICIS (National Integrated Cyber Infrastructure System) e-research grant QICSA is kindly acknowledged.

\nocite{*}
\bibliography{main}

\pagebreak
\section*{\label{sec:appendix}Appendix}
The architecture of the convolutional neural network model developed in this study is summarized in \ref{fig:fig3} below. Here the number of input features is three and the number of outputs is two. This model can be adapted for systems of varying complexity as seen in this work. The number of input features for an N-level system is $2(N)-1$ and the number of outputs is $2(N-1)$.
\begin{figure}[h]
    \centering
    \includegraphics[height=14cm]{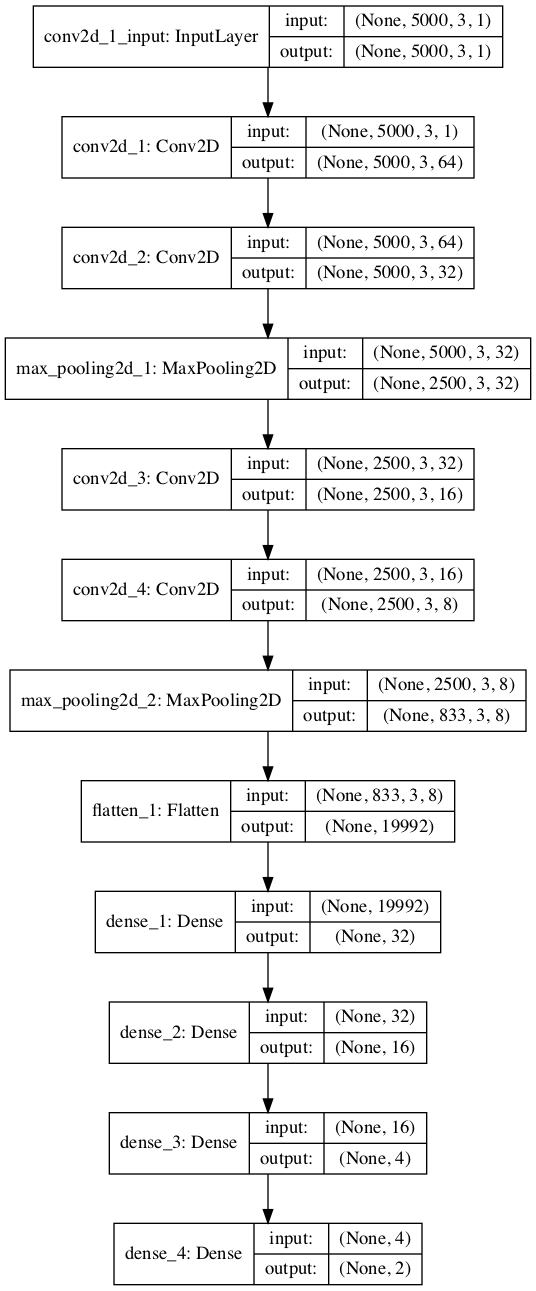}
    \caption{In this figure, the architecture of the convolutional neural network suited for the prediction of parameters of two-level system Hamiltonians is shown. The input/ output parameters on the right hand side of the figure describe the dimensions of the data going into and out of each layer. The left hand side describes the type of layer used in the network where the following key can be used - conv2d: 2D convolution layer, max\_pooling2d: max pooling layer for 2D inputs, flatten: layer to flatten the input and dense: deeply connected layer.}
    \label{fig:fig3}
\end{figure}
\clearpage
\end{document}